\documentclass[%
  aps,             
  prf,        
  superscriptaddress, 
  floatfix         
]{revtex4-2}
\usepackage{physics}
\usepackage{amssymb}
\usepackage{bm}
\usepackage{graphicx}
\usepackage{xcolor}
\usepackage{subcaption}
\usepackage{siunitx}
\usepackage{tikz}
\usepackage{booktabs}
\usepackage{multirow}

\newcommand{\mean}[1]{\left\langle #1 \right\rangle}

\begin{document}

\title{Imprints of turbulence on heterogeneous deposition of adhesive particles}

\author{Max Herzog, Jesse Capecelatro}
\date{July 2025}%


\begin{abstract}
We present direct numerical simulations (DNS) of particle deposition in a turbulent channel flow, incorporating a viscoelastic soft-sphere collision model with temperature-dependent van der Waals adhesion. Particle-wall contact is governed by an adhesion number that varies with temperature, enabling exploration of a wide range of deposition behaviors. Deposition is strongly heterogeneous, especially for inertial particles, where rolling and sliding enhance nonuniformity. Spanwise radial distribution functions reveal that deposited particles form streaks with characteristic spacing set by near-wall two-point velocity correlations. A clustering metric confirms that high-inertia, low-adhesion particles deposit in elongated, anisotropic patterns due to spanwise migration driven by velocity fluctuations. Finally, it is shown that this heterogeneity in deposition leads to localized wall wear rates exceeding ten times the mean, with the most severe wear associated with particles that are carried to the wall in clusters by sweep events.
\end{abstract}

\maketitle


\section{Introduction}
Particle-turbulence interactions play a central role in erosion, fouling, and deposition across a wide range of engineering and environmental systems. These effects are especially critical in wall-bounded flows, where solid particles interact with near-wall turbulent structures before depositing on surfaces. Such flows are encountered in diverse applications, including fouling in heat exchangers \cite{Kapustenko2023-ds}, particle deposition in combustion systems \cite{Kleinhans2018-un}, and water droplet accumulation and icing on aircraft surfaces \cite{Yamazaki2021-jg}. A particularly challenging example is the ingestion of dust and sand in gas turbine engines, where next-generation turbomachinery operates at temperatures near or above the melting point of entrained particles--conditions that promote deposition and accelerate erosion and fouling of hot-section components \cite{Nieto2021-pt, Levi2012-bv}.

\begin{figure}[bht]
    \centering
    \includegraphics[width=1.0\linewidth]{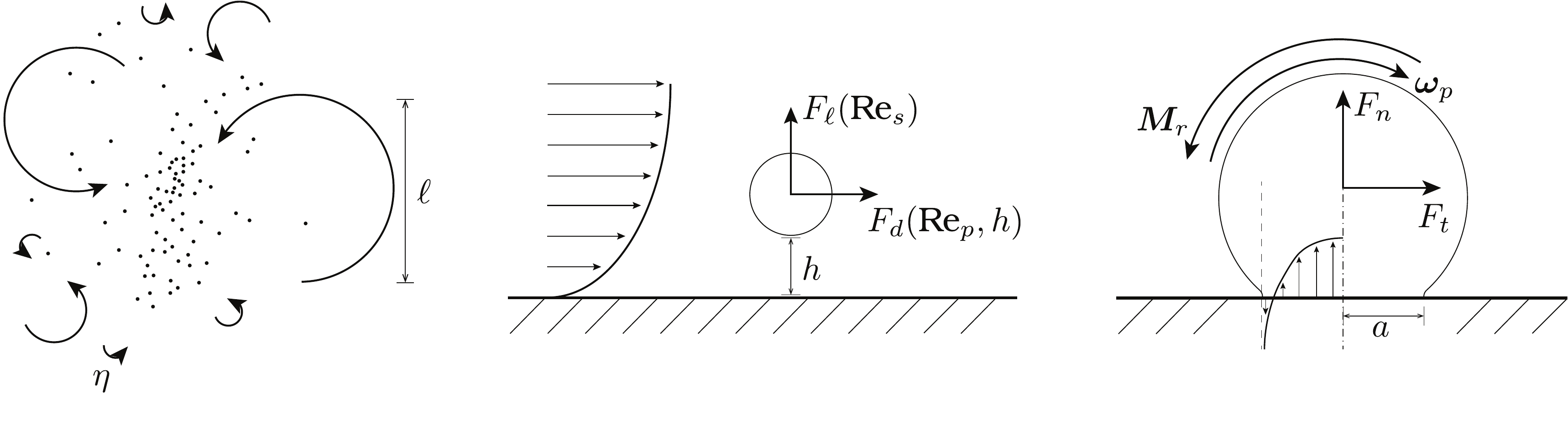}
    \caption{A schematic depicting the multi-scale nature of particle deposition, beginning with preferential concentration in the free-stream dictated by the integral, $\ell$, and Kolmogorov, $\eta$, length scales, fluid-mediated particle-wall interactions in the near-wall region that depend on the distance between the particle and the wall, $h$, down to the sub-particle diameter with particle-wall contact patch radius, $a$.}
    \label{fig:partLife}
\end{figure}

Particle deposition in wall-bounded turbulence arises from interactions spanning a broad range of spatial and temporal scales. As illustrated in Fig.~\ref{fig:partLife}, three key regimes can be identified: (i) in the outer flow, particles are transported by turbulence and preferentially concentrate \cite{reeks1983transport, eatonPreferentialConcentrationParticles1994}; (ii) in the near-wall region, particles are transported to the surface by coherent structures in the flow and experience modified drag and lift forces \cite{shiLiftForcesSolid2020, mclaughlinAerosolParticleDeposition1989}; and (iii) upon surface contact, particles undergo viscoelastic and adhesive interactions that govern rebound, sticking, sliding, and rolling behavior \cite{marshallAdhesiveParticleFlow2014}. The disparate nature of these multiscale processes poses significant challenges for predictive modeling. The present work introduces a numerical framework that captures these coupled effects and quantifies the resulting heterogeneous deposition patterns and surface wear.

A hallmark of particle-laden flows is that particles with finite mass do not follow fluid streamlines. This leads to preferential concentration, where particles are centrifuged out of turbulent eddies and accumulate in regions of high strain and low vorticity \cite{Squires1991-jd}. The degree to which particles respond to turbulent fluctuations is characterized by the Stokes number, the ratio of the particle response time to a relevant fluid time scale. Preferential concentration is most pronounced at intermediate Stokes numbers, where particles are responsive to small-scale turbulence but retain sufficient inertia to decouple from the flow. This intermediate regime is the primary focus of the present study.

Preferential concentration is fundamentally a two-point phenomenon; particle pairs experience spatially-correlated turbulent fluctuations that drives them to concentrate. This was identified in the collective works by \citet{zaichikPairDispersionPreferential2003, zaichikStatisticalModelsPredicting2009}, whose theory pointed to two-point velocity moments as a source of preferential concentration: particle pairs with finite Stokes number accumulate as they migrate away from regions of high turbulence intensity. This mechanism bears strong similarity to turbophoresis in inhomogeneous, wall-bounded flows \cite{reeks1983transport}, where particles drift from more turbulent regions toward the wall. This accumulation defines the second regime in Fig.~\ref{fig:partLife}, where particles interact with coherent structures and are carried to (from) the wall by sweep (ejection) events \cite{marchioli2002, narayanan2003, Soldati2009-wl}. In such events, low-speed fluid is lifted away from the wall (ejection), and high-speed fluid moves down and replaces it (sweep) \cite{pope2000}. Particles respond to these structures according to their friction Stokes number,  ${\rm{St}}^+$. Particles with ${\rm{St}}^+\in[1,50]$ are strongly influenced by near-wall structures and exhibit the most heterogeneity. Those with ${\rm{St}}^+\sim10$ resonate with the coherent structures, forming elongated streak-like patterns in the flow \cite{Soldati2009-wl, narayanan2003}. Particles with response times comparable to sweep timescales tend to reach the wall and remain there, while low-inertia particles (${\rm{St}}^+\lesssim10$) are re-entrained by ejections. Conversely, particles with ${\rm{St}}^+\gg10$ carry memory of outer-layer turbulence and impact the wall ballistically. The range ${\rm{St}}^+\in[10,100]$ spans these behaviors and is particularly relevant to dust ingestion in gas turbine engines \cite{diaz-lopez2025}. While electrostatic forces may also influence deposition in this regime \cite{yaoMultifidelityUncertaintyQuantification2022, ruan2025, guha2008}, we restrict our focus to particle-turbulence interactions and neglect electrostatics in the present study.

The third panel in Fig.~\ref{fig:partLife} focuses on particle-wall contact mechanics. Particles that reach the surface may stick, rebound, slide, or roll. Elastic, plastic, viscous, and adhesive forces depend on particle material properties and impact velocities. Two classes of models exist for modeling particle-wall collisions. The soft-sphere approach treats particles as adhesive springs, determining collisional forces as a function of particle deformation. Several models that include viscous, elastic, plastic, and adhesive forces have been proposed \cite{Marshall2009-tv, thorntonTheoreticalModelStick1998, Gaskell2022-qe}. An advantage of soft-sphere models is that they resolve tangential and rotational forces during contact, capturing rolling and sliding behavior. Soft-sphere models also allow for sustained many-body contact. However, such models require small time s-eps-converted-to.pdf in order to resolve particle deformation. This is exacerbated by the high spring stiffness imposed when using realistic material properties. Hard-sphere models avoid this cost by assuming instantaneous particle collisions~\cite{hoomans1996discrete}, whereby collisional forces and moments are treated as impulses. Consequently, capturing particle rolling and sliding is not trivial. 

Particle adhesion arises from several mechanisms, including van der Waals forces, liquid bridging, electric double-layer interactions, and sintering \cite{marshallAdhesiveParticleFlow2014}. Experiments consistently report increased sticking at elevated temperatures \cite{berbnerInfluenceHighTemperatures1994, Plewacki2020-zj, bons2022, reagle2014study, delimont2015effect}, attributed to mechanisms such as enhanced plasticity, sintering, liquid bridging, and changes in surface chemistry. Additional adhesion mechanisms, such as particle-material shock, particle-particle hammering, and surface fracture, become important at high impact velocities~\cite{singh2021, klinkov2005}. Adhesion is highly sensitive to material structure, surrounding medium, surface conditions, and temperature \cite{bergstrom1996, bergstrom1997hamaker, pinchuk2016}, complicating the definition of robust material parameters for modeling. Consequently, many studies rely on tuning adhesion models and material properties to match experimental data~\cite{Bons2017-gp, singhParticleDepositionModel2015, yu2017}. While effective for specific conditions, this approach entangles material properties with model construction, limiting predictive capability and generalizability.

While considerable effort has been devoted to modeling particle deposition under various flow conditions and configurations \cite{Gaskell2022-qe, Bons2017-gp, Whitaker2016-vc, Borello2016-db, Suman2019-eo, Wolff2018-ew}, fewer studies have investigated the \textit{spatial structure} of deposited particles and its relationship to turbulence. Using direct numerical simulation (DNS) of a turbulent channel, \citet{narayanan2003} examined deposition mechanisms and observed streak-like deposit patterns. However, their simulations omitted near-wall contact physics, such as adhesion, rebound, sliding, and rolling, as illustrated in the third panel of Fig.~\ref{fig:partLife}. More recent efforts that do incorporate particle-wall contact often rely on Reynolds-averaged Navier–Stokes (RANS) simulations combined with stochastic subgrid-scale models for unresolved velocity fluctuations ``seen'' by the particles~\cite{Gaskell2022-qe, Vadgama2021-pu}. Subgrid-scale models most commonly solve for one-point statistical moments, which are insufficient in capturing the two-point moments that drive particle heterogeneity \cite{Marchioli2017-tf}. Experimental studies also highlight turbulence-driven spatial variability in deposition and wear. For instance, \citet{molina2019application, molina2020erosion} identified patterns of abrasive wear linked to turbulence–particle interactions, though their findings were limited to a priori surface analysis without direct resolution of flow structures. 

This work employs DNS of a turbulent channel flow within an Euler--Lagrange framework to quantify heterogeneous particle deposition and its underlying mechanisms, emphasizing the roles of turbulence, particle inertia, and adhesion strength. Particle collisions are modeled using a viscoelastic soft-sphere approach incorporating van der Waals adhesion. The paper is organized as follows: Section~\ref{sec:physParam} details the physical parameters used; Section~\ref{sec:compApproach} describes the computational models and methods; Section~\ref{sec:results} presents validation of the framework, analyzes particle heterogeneity and its origins, and examines its effect on abrasive wear; finally, Section~\ref{sec:conclusion} summarizes the findings and conclusions.


\section{Simulation details}\label{sec:caseDesc}

\subsection{Configuration}\label{sec:physParam}

\begin{figure}
    \centering
    \includegraphics[width=0.95\linewidth]{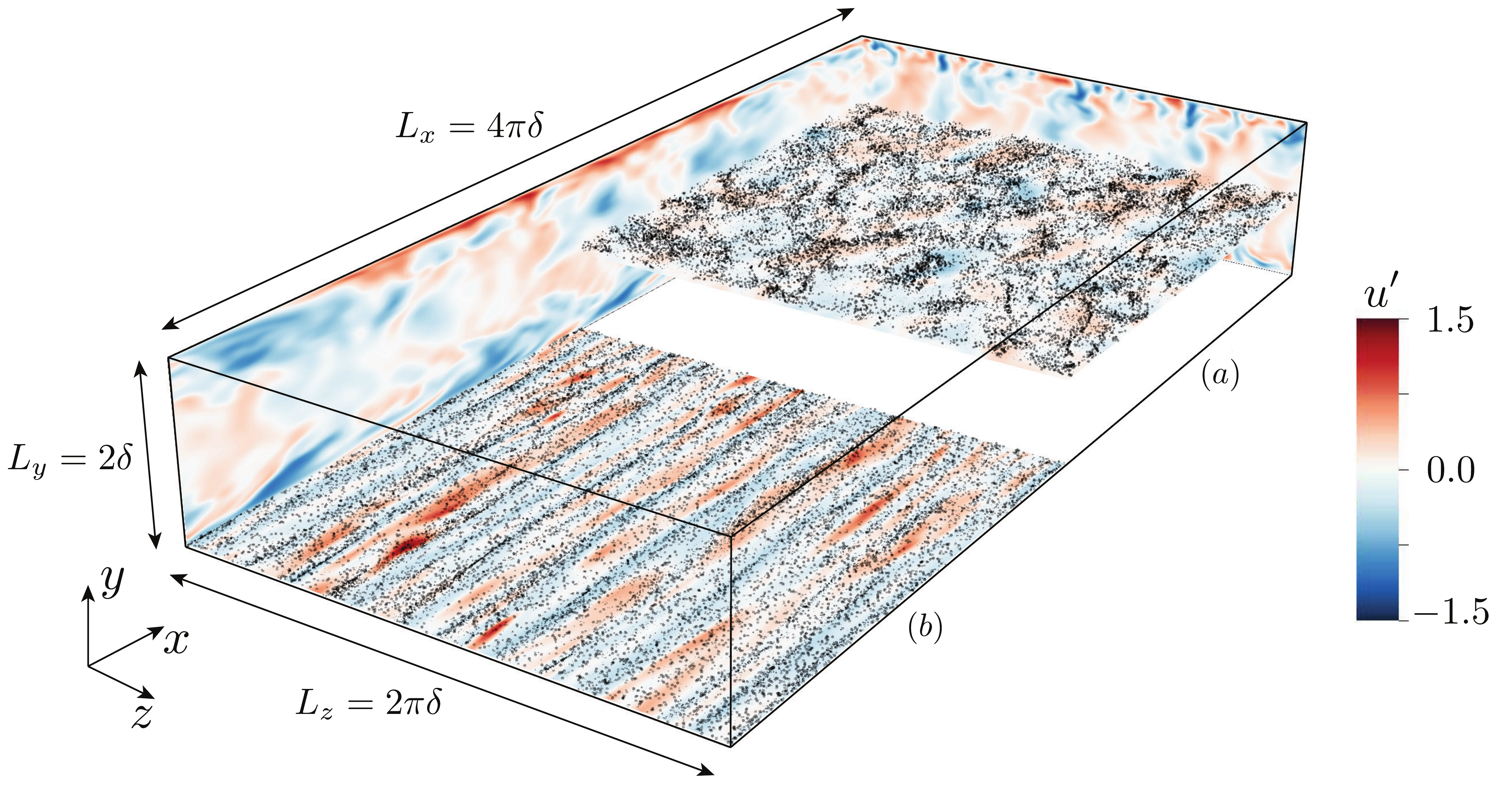}
    \caption{Particle positions overlaid on contours of the streamwise velocity fluctuations in (a) the centerline and (b) a region close to the wall ($y^+\in[2,8]$), highlighting particle clustering in low-speed streaks in the near-wall region and preferential concentration in the free-stream. Particles are polydisperse with ${\rm{St}}^+\in[0.1, 1,10,30]$. The friction Reynolds number is $\rm{Re}_\tau=200$.}
    \label{fig:channelSegment}
\end{figure}

Figure~\ref{fig:channelSegment} shows the turbulent channel flow considered in this study. The flow is oriented such that the streamwise, wall-normal, and spanwise directions correspond to $x$, $y$, and $z$, respectively. Periodic boundary conditions are imposed in the streamwise and spanwise directions. Wall roughness is not accounted for. The channel dimensions are $L_x\times L_y\times L_z=4\pi\delta\times2\delta\times2\pi\delta$, with $\delta=\SI{1.5}{\centi\meter}$. Air is taken as the working fluid ($\mu=\SI{1.8e-5}{\kilogram\per\meter\per\second}$, $\rho=\SI{1.2}{\kilogram\per\meter\cubed}$). The bulk Reynolds number is ${\rm{Re}}_b=2\delta\overline{u}/\nu=6355$ where $\overline{u}=\int_0^\delta\mean{u}\dd{y}/\delta$ is the bulk velocity. The angle brackets $\mean{\cdot}$ denote a quantity averaged in time and in the homogeneous directions. The friction Reynolds number is ${\rm{Re}}_\tau=u_\tau\delta/\nu=200$, where the friction velocity is given by
\begin{equation}
    u_\tau=\sqrt{\nu\left.\dv{\mean{u}}{y}\right\vert_{y=0}}.
\end{equation}
The friction length scale is $\delta_\nu=\nu/u_\tau$ and the friction time scale is $\tau_\nu=\nu/u_\tau^2$. These wall units are used for non-dimensionalization throughout this paper, denoted by a superscript $+$.   

The friction Stokes number is given by ${\rm{St}}^+=\tau_{p}/\tau_\nu$ where $\tau_{p}=\rho_p d_p^2 / (18\mu)$. The particle diameter is varies such that the Stokes numbers ${\rm{St}}^+\in[0.1, 1, 10, 30]$. The lower two Stokes numbers are used for validation of the computational framework. Deposition rates for ${\rm{St}}^+=0.1$ and $1$ are low and are not carried to the wall by their inertia \cite{marchioli2002, narayanan2003}. Instead, we focus on ${\rm{St}}^+=10$ and $30$ for the majority of the results in this paper. These Stokes numbers are more relevant to heterogeneous deposition and wear because they have timescales comparable to the near-wall coherent structures, exhibit high deposition rates, and can reach the wall by impaction instead of diffusion. 

Visco-elastic-adhesive contact mechanics are dictated by the particle diameter $d_p$, the particle density $\rho_{p}$, the Young's modulus $E$, the Poisson ratio $\nu_{p}$, the viscous coefficient of restitution $e_0$, the coefficient of friction $\mu_f$, and the surface free energy $\gamma$. While some of these properties are known as a function of temperature, others are not readily available. The free surface energy is largely extrinsic and is consequently difficult to characterize. In place of varying material properties as a function of temperature, we instead approximate the effect of temperature-dependent adhesion by varying the adhesion number, defined in this study as
\begin{equation}
   {\rm{Ad}}^+=\frac{2\gamma}{\rho_p d_p u_\tau^2}.
\end{equation}
An adhesion number of $\rm{Ad}^+ = 0$ corresponds to non-adhesive (purely elastic) particles, while $\rm{Ad}^+ \to \infty$ represents fully adhesive particles that stick upon contact (all-stick). The adhesion number is most accurately defined using the particle velocity immediately prior to surface impingement \cite{chenExponentialScalingEarlystage2019}. However, this velocity is not readily available in many flow configurations. For this reason, we define $\rm{Ad}^+$ using characteristic wall turbulence scales. Our goal is to establish a connection between particle adhesion behavior and more accessible wall-bounded turbulence parameters, such as the friction Reynolds number, $\rm{Re}_\tau$.

A summary of relevant material properties considered in this study is provided in Table~\ref{tab:partProp}. All non-adhesive properties are set to be consistent with crystalline quartz particles impacting a 310 stainless steel wall at room temperature. Crystalline quartz comprises the majority of atmospheric sand and dust globally~\cite{Nieto2021-pt}. The Hamaker constant, $A$, is related to $\gamma$ via $\gamma=A/24\pi\delta_{\rm{gap}}^2$ where $\delta_{\rm{gap}}=\SI{0.165}{\nano\meter}$ is the equilibrium gap thickness \cite{marshallAdhesiveParticleFlow2014}. The range of Hamaker constants used here lies within a realistic range of values for other common atmospheric particulates \cite{yao2022, bergstrom1997hamaker}. 

\begin{table}[]
    \centering
    \caption{Particle parameters considered in this study. For all cases, the Young's modulus $E=63.5$ [GPa], particle density $\rho_p=2625$ [$\SI{}{\kilogram\per\meter\cubed}$], Poisson ratio $\nu_p=0.054$, coefficient of friction $\mu_f=0.55$, and viscous coefficient of restitution $e_0=0.9$.}
    \begin{tabular}{ccccc}
        \toprule
        $\rm{Ad}^+$ & $\rm{St}^+$ & $d_p$ [$\SI{}{\micro\meter}$] & $\gamma$ [$\SI{}{\newton\per\meter}$] & $A$ [$\SI{}{J}$]  \\
        \midrule
        \multirow{2}{*}{15}  
          & 10 & 23.5   & 0.018 & $\SI{3.799e-20}{}$   \\ 
          & 30 & 40.704 & 0.032 & $\SI{6.580e-20}{}$   \\ 
        \midrule
        \multirow{2}{*}{50}  
          & 10 & 23.5   & 0.062 & $\SI{12.663e-20}{}$  \\ 
          & 30 & 40.704 & 0.106 & $\SI{21.933e-20}{}$  \\ 
        \midrule
        \multirow{2}{*}{150} 
          & 10 & 23.5   & 0.185 & $\SI{37.988e-20}{}$  \\ 
          & 30 & 40.704 & 0.321 & $\SI{65.799e-20}{}$  \\ 
        \bottomrule
    \end{tabular}
    \label{tab:partProp}
\end{table}


\subsection{Computational approach}\label{sec:compApproach}
\subsubsection{Fluid}
The gas phase is described by the incompressible, constant-density Navier--Stokes equations given by
\begin{equation}
\pdv{\bm{u}}{t} + (\bm{u}\cdot\nabla)\bm{u} = -\frac{1}{\rho}\nabla p + \nu \nabla^2 \bm{u} \quad {\rm{and}} \quad \nabla\cdot\bm{u}=0,
\end{equation}
where $\bm{u}$ is the fluid velocity, $\rho$ is the fluid density, $p$ is the pressure, and $\nu=\mu/\rho$ is the kinematic viscosity. All simulations were performed using the open-source flow solver NGA2 \cite{DesjardinsUnknown-et, desjardins2008high, Capecelatro2013-gf}. The Navier--Stokes equations are solved using a second-order finite volume scheme on a staggered grid. The equations are advanced using a second-order semi-implicit Crank--Nicholson scheme. The pressure Poisson equation is solved to machine precision via FFTW. The domain is discretized with $N_x\times N_y\times N_z=192^3$ grid points and uniform mesh spacing in $x$ and $z$. Wall-normal grid points are concentrated near the wall according to
\begin{equation}
y_i=\frac{L_y\tanh[2\alpha_y(i-1)/N_y-1]}{\tanh[\alpha_y]},
    \label{eq:grid}
\end{equation}
where $\alpha_y=2.1$ is the stretching parameter and $i$ is the grid index. The non-dimensional grid spacing is $\Delta x^+=13.09$, $\Delta y^+\in[0.268,4.507]$, and $\Delta z^+=6.55$.

We note that the flows under consideration here are dilute in solid volume fraction ($\sim\mathcal{O}(10^{-5})$). Consequently, the effects of inter-phase momentum transfer and inter-particle interactions on the gas and disperse phase are neglected. While the long-time accumulation of particles on bounding surfaces can lead to significant on-wall agglomeration, we leave the results and consequences of particle-particle interaction and pileup available for future study. 

\begin{figure}
    \centering
    \begin{subfigure}[b]{0.49\textwidth}
      \centering
      \includegraphics[width=\textwidth]{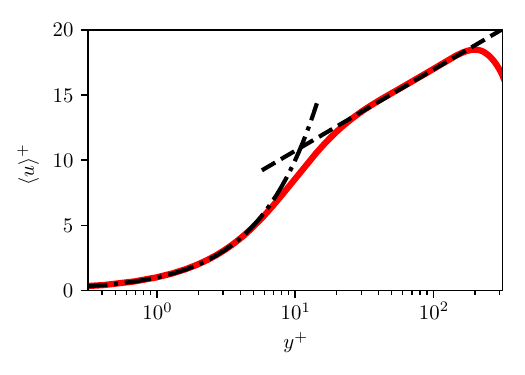}
      \caption{Mean velocity.}
    \end{subfigure}
    \begin{subfigure}[b]{0.49\textwidth}
      \centering
      \includegraphics[width=\textwidth]{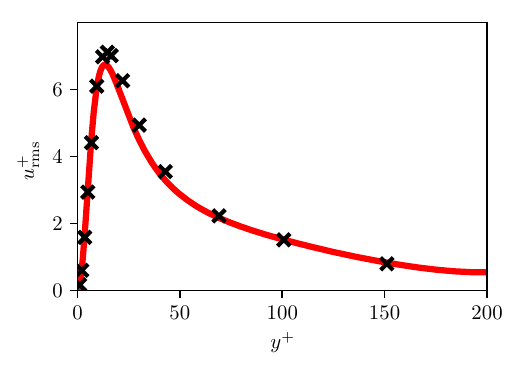}
      \caption{RMS velocity.}
      \label{fig:sub4}
    \end{subfigure}
    \caption{Single phase streamwise velocity statistics. Lines in (a) are linear and log-law relations and symbols in (b) are from the DNS of \cite{LeeMoser2015}}
    \label{fig:channelVal}
\end{figure}

Single-phase simulations were run until a steady state was reached.Validation of the streamwise velocity statistics is shown in Fig.~\ref{fig:channelVal}. Overall, the results show excellent agreement with classical turbulence theory and prior DNS data for both the mean and root-mean-square (RMS) velocity profiles.

\subsubsection{Particles}
Particles are advanced according to
\begin{equation}
m\dv{\bm{v}}{t}=\bm{F}_d+\bm{F}_\ell+\bm{F}_{b}+\bm{F}_{\rm{col}}, \quad I\dv{\bm{\omega}_p}{t}=\bm{M}_{\rm{col}},
\end{equation}
where $m$ is the particle mass, $\bm{v}$ is the particle velocity, $I=md_p^2/10$ is the moment of inertia, and $\bm{\omega}_p$ is the angular velocity of the particle. Drag, lift, and Brownian forces, given by $\bm{F}_d$, $\bm{F}_\ell$, and $\bm{F}_b$, respectively, comprise the fluid coupling. $\bm{F}_{\rm{col}}$ and $\bm{M}_{\rm{col}}$ represent collision forces and torques. 

Particles near a wall experience modified lift and drag forces. Decades of work have been devoted to developing near-wall lift corrections \cite{shiLiftForcesSolid2020}, however, most models are complex in their construction and are valid only in specific regimes. Furthermore, near-wall lift forces do not play an important role for $d_p^+<1$ \cite{gao2024relevance}. The largest particles considered in this work are $d_p^+\approx0.5$. We thus do not employ a wall-corrected lift model and instead use the Saffman model \cite{saffman1965lift}, which has demonstrated adequate performance in channel flows \cite{costa2020interface}. The lift force is given by
\begin{equation}
\bm{F}_\ell = 1.615\mu d_p ||\bm{v}-\bm{u}|| \sqrt{\frac{d_p^2||\bm{\omega}||}{\nu}} \frac{\bm{\omega}\times(\bm{v}-\bm{u})}{||\bm{\omega}||\,||\bm{v}-\bm{u}||}
\end{equation}
 
The near-wall drag force is given by \cite{Zeng2009-qj}
\begin{equation}
    \bm{F}_d=\frac{C_d}{\tau_{p}}(\bm{u} - \bm{v}), \quad C_d=f(h)\left(1 + \alpha_s(h)\text{Re}_p^{\beta_s(h)}\right),
    \label{eq:zengDrag}
\end{equation}
where $h$ is the non-dimensional distance between the particle and wall (in particle diameters) and ${\rm{Re}}_p$ is the particle Reynolds number, given by
\begin{equation}
     {\rm{Re}}_p=\frac{\rho||\bm{v}-\bm{u}||d_p}{\mu}.
\end{equation}
We note that \eqref{eq:zengDrag} tends to the classical Schiller--Naumann drag correction away from the wall.

The deposition of inertial particles with ${\rm{St}}^+<0.2$ is dominated by turbulent diffusion and Brownian forces \cite{young1997theory}. We model Brownian dynamics as a stochastic force in the particle equations of motion, given by
\begin{equation}
    \bm{F}_b = \sqrt{\frac{mk_BT_f}{\tau_{p}}}\dd{\bm{W}},
\end{equation}
where $\dd{\bm{W}}\sim\mathcal{N}\left(0,\sqrt{\dd{t}}\right)$ is an increment of the Wiener process, $T_f$ is the fluid temperature in units Kelvin, and $k_B$ is the Boltzmann constant. 

The particle equations of motion are advanced using a modified Heun method \cite{roberts2012modify}, which provides first-order convergence when the Brownian force dominates and second-order accuracy in the limit $\bm{F}_b=0$. This maintains near second-order accuracy when the noise component is small compared to the hydrodynamics, as is the case for the ${\rm{St}}^+=10$ and $30$ particles we focus on in this study.

\subsubsection{Particle-wall contact}\label{sec:ssmodel}
Particle-wall collisions are modeled using the soft-sphere model presented in \cite{Marshall2009-tv}, in which particles are allowed to deform and collisions occur over a finite duration. Contact forces are computed based on the particle overlap in the normal and tangential directions, $\delta_{n/t}$. The model incorporates viscoelastic dissipation, Hertzian elastic rebound, and van der Waals adhesion, following the Johnson--Kendall--Roberts (JKR) theory \cite{Johnson1971-yo}. The strength of adhesive interactions relative to elasticity is characterized by the Tabor parameter \cite{tabor1977}, defined as $\lambda_T = (2d_p\gamma^2 / E^2\delta_{\rm{gap}}^3)^{1/3}$. The JKR model is valid in the regime $\lambda_T \gtrsim 10$, where particle deformation produces a flattened contact patch across which adhesion acts \cite{marshallAdhesiveParticleFlow2014}. In our simulations, the minimum Tabor parameter is approximately $\lambda_T \approx 15$, confirming the applicability of the JKR theory.

Tangential forces are accounted for using a spring-dashpot force that saturates at the Coulomb limit. The contact force is written as the sum of the normal and tangential forces, according to
\begin{equation}
    \bm{F}_{\rm{col}} = F_n\bm{n} + F_t\bm{t},
\end{equation}
where $\bm{n}$ and $\bm{t}$ are the normal and tangential unit vectors, respectively. The normal force includes elastic-adhesive and dissipative forces, given by
\begin{equation}
    F_n = F_{ne} + F_{nd},
\end{equation}
with
\begin{equation}
    F_{ne} = 4F_C\left[\left(\frac{a}{a_0}\right)^3 - \left(\frac{a}{a_0}\right)^{3/2} \right],
\end{equation}
where $F_{nd}=-(\bm{v}\cdot\bm{n})\eta_n$ and $\eta_n=2\alpha\sqrt{mEa/3}$ is the dissipation coefficient, $a$ is the contact radius, and $\alpha$ is determined from \citet{Marshall2009-tv}.

The non-dimensional contact radius, $a/a_0$, is found as the root of
\begin{equation}
    \frac{\delta_n}{\delta_C}=6^{1/3}\left[ 2\left(\frac{a}{a_0}\right)^2 - \frac{4}{3}\left(\frac{a}{a_0}\right)^{1/2} \right],
\end{equation}
where $a_0$ is the equilibrium contact radius, $F_C$ is the critical force at separation, and $\delta_C$ is the inter-particle distance at which separation occurs. These quantities are determined from the particle geometric and material properties according to
\begin{equation}
    a_0=\left(\frac{9\pi\gamma d_p^2}{4E}\right)^{1/3}, \quad F_C=\frac{3}{2}\pi\gamma d_p, \quad \textrm{and} \quad \delta_C = \frac{a_0^2}{6^{1/3}d_p}.
\end{equation}

The tangential component of the contact force is
\begin{equation}
    F_t = \min(|k_t\bm{\delta_t} - \bm{v}_{r}\eta_t|, \mu_f|F_{ne}+2F_C|),
    \label{eq:tancol}
\end{equation}
with the direction of the force $\bm{t}$ lying in the same direction as the tangential overlap $\bm{\delta}_t=\int_{t_0}^t \bm{v}_r \dd{t'}$, $\eta_t=\eta_n$, and $k_t=8Ga$ with $G=E/2(1+\nu_p)$ the shear modulus. The tangential relative velocity at contact for a particle colliding with a wall is
\begin{equation}
    \bm{v}_r = \bm{v} - (\bm{v}\cdot\bm{n})\bm{n}-\frac{d_p}{2}\bm{n}\times\bm{\omega}_p,
    \label{eq:relvel}
\end{equation}
where the normal vector $\bm{n}$ is defined positive into the wall. 

Adhesive van der Waals forces introduce a mechanism for rolling resistance during particle-particle and particle-wall contact. In the case of a particle impacting a stationary wall, the rolling displacement moment arm is determined by
\begin{equation}
    \bm{\xi}=\frac{1}{2}\int_{t_0}^{t}d_p\bm{n}\times\bm{\omega}_p\dd{t'}.
\end{equation}
The rolling displacement saturates at the critical value $\xi_{\rm{crit}}=\theta_{\rm{crit}}d_p/2$, where $\theta_{\rm{crit}}=\SI{40e-3}{}$ \cite{Marshall2009-tv}, at which point the displacement vector is set to $\bm{\xi}^{\rm{new}}=\xi_{\rm{crit}}\bm{\xi}^{\rm{old}}/||\bm{\xi}^{\rm{old}}||$. The rolling adhesion is then determined by
\begin{equation}
    \bm{M}_r=2F_C\left(\frac{a}{a_0}\right)^{3/2}d_p\bm{n}\times \bm{\xi}.
    \label{eq:rollmom}
\end{equation}
The total collisional moment is the sum of the rolling resistance and the moment generated by the tangential forces given by
\begin{equation}
    \bm{M}_{\rm{col}}=\bm{M}_r+\frac{d_p}{2}\bm{n}\times(F_t\bm{t}).
\end{equation}

The contact mechanics of inter-particle and particle-wall collisions occur on small timescales. This can lead to prohibitively small time step constraints when running with a large number of particles or if the goal in using the model is to study phenomena which occur on larger time scales. It is common to artificially soften the particle properties to permit larger time s-eps-converted-to.pdf. We use the method outlined by \citet{Haervig2017-pb}, where $E$ is artificially reduced to a lower value, $E_{\rm{mod}}$, and $\gamma_{\rm{mod}}=\gamma\,(E_{\rm{mod}}/E)^{2/5}$. Reducing the stiffness and surface free energy leads to a natural reduction in the rolling resistance. To combat this, only the ratio $(a/a_0)^{3/2}$ in Eq.~\eqref{eq:rollmom} is modified when a reduced particle stiffness is used. In accordance with \citet{Haervig2017-pb}, all other terms are left unaltered. In this work, all simulations were conducted with a softening ratio $E/E_{\rm{mod}}=1000$. Validation of the soft-sphere model is provided in Appendix~\ref{sec:appendix}.

We also utilize a domain segregated particle sub-stepping scheme to further reduce the computational burden associated with the micro-scale physics of the particle-wall collisions. In this framework, the fluid advances with a timestep restricted by the CFL, $\Delta t_f=\min(\Delta_i/|u_i|)$, where the subscript $i$ is a component index.  Each particle evolves independently with a time step determined by the wall-normal position, according to
\begin{equation}
    \Delta t_p=
    \begin{cases}
        \min(\Delta_i/|v_i|) & \textrm{if}\quad  y \geq 5\,d_p  \\
        0.1d_p/||\bm{v}||    & \textrm{if}\quad  0 < y < 5\,d_p \\
        \tau_{\rm{col}}/50   & \textrm{if}\quad  y = 0,          \\
    \end{cases}
\end{equation}
where the collision timescale is estimated as the Rayleigh wave velocity given by \cite{burns2019}
\begin{equation}
    \tau_{\rm{col}}=\frac{\pi d_p}{2\beta}\sqrt{\frac{\rho_p}{G}} \quad {\rm{and}} \quad \beta=0.8766+0.163\nu_p.
    \label{eq:colTime}
\end{equation}

After steady state is reached, $\SI{5e5}{}$ particles of each diameter were initialized randomly in the channel. We note that particles were not re-injected upon depositing, as is sometimes done. The simulation was then run to steady state with purely elastic collisions. As in \cite{marchioliStatisticsParticleDispersion2008}, the particle phase was determined to be at steady state once the near-wall concentration of particles reached a constant value. Each Stokes number case then had its own initial conditions that were identical for every adhesion number run. Once a statistical steady state was reached, the soft-sphere collision model was enabled. All adhesion numbers for a given ${\rm{St}}^+$ have the same initial conditions. Simulations were run until at least 30\% of particles had deposited on the wall. We note that running the simulations until all particles deposit is not computationally feasible due to the low deposition and adhesion rates.


\section{Results}\label{sec:results}

\subsection{Particle statistics}

\begin{figure}
  \centering
  \begin{subfigure}[t]{0.32\textwidth}
    \centering
    \includegraphics[width=\textwidth]{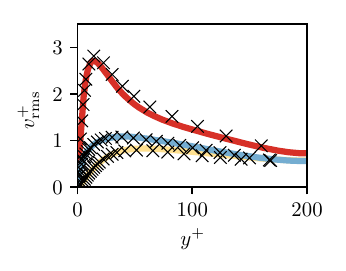}
    \caption{${\rm{St}}^+=1$}
    \label{fig:sub1}
  \end{subfigure}
  \begin{subfigure}[t]{0.32\textwidth}
    \centering
    \includegraphics[width=\textwidth]{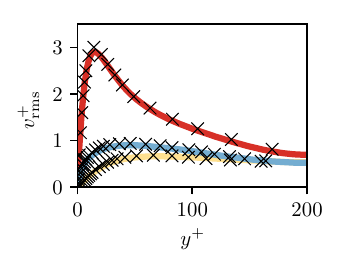}
    \caption{${\rm{St}}^+=10$}
    \label{fig:sub2}
  \end{subfigure}
  \begin{subfigure}[t]{0.32\textwidth}
    \centering
    \includegraphics[width=\textwidth]{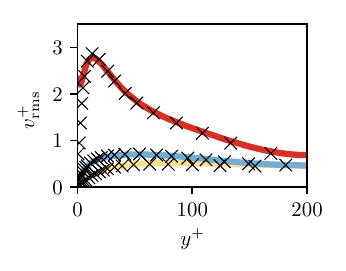}
    \caption{${\rm{St}}^+=30$}
  \end{subfigure}
  \caption{Particle velocity RMS profiles computed for three Stokes numbers. Symbols are from the DNS of \citet{arcen2006influence}.}
  \label{fig:pRMSval}
\end{figure}

Figure~\ref{fig:pRMSval} shows RMS velocity profiles for the particle phase at three Stokes numbers, ${\rm{St}}^+\in[1,10,30]$. Symbols represent results from the DNS by Arcen \cite{arcen2006influence}. Overall excellent agreement is observed. A small discrepancy appears at low $y^+$ in the streamwise RMS velocity for ${\rm{St}}^+=30$, which we attribute to the drag model we use in this study. This observation aligns with prior findings~\cite{jin2016}.

\begin{figure}
    \centering
    \includegraphics[width=0.5\linewidth]{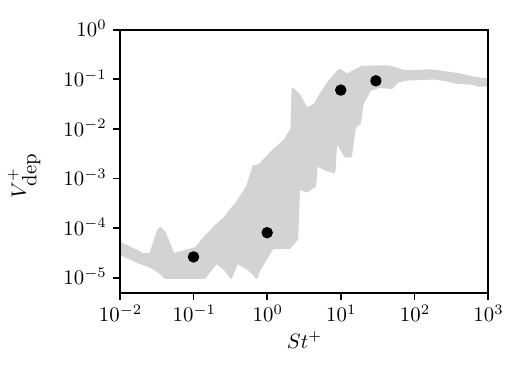}
    \caption{Validation of non-dimensional deposition velocity against the literature. Markers denote the results of this study (${\rm{St}}^+\in[0.1,1,10,30]$) with $\rm{Ad}^+=\infty$. The shaded region covers data from \cite{sehmel1968aerosol, Liu1974-qh, Schwendiman1962-hs, Friedlander1957-yv}.}
    \label{fig:depvel}
\end{figure}

The non-dimensional deposition velocity, $V^+_{\rm dep}$, is computed as~\cite{kallio1989numerical}
\begin{equation}
    V^+_{\rm{dep}}=\frac{J_w}{\rho_{pw}u_\tau},
\end{equation}
where $J_w$ is the particle mass flux at the wall, and $\rho_{pw}$ is the mass of deposited particles per unit volume. Figure~\ref{fig:depvel} compares $V^+_{\rm dep}$ from our simulations with ${\rm Ad}^+=\infty$ (all-stick) across a range of ${\rm St}^+$ values with available experimental data. Our results fall within the bounds of reported measurements. Three distinct deposition regimes are captured in this Stokes number range: (i) the diffusion dominated regime for ${\rm{St}}^+<1$, which exhibits low deposition rates ($\mathcal{O}(10^{-4})-\mathcal{O}(10^{-5})$); (ii) the transition region ($1<{\rm{St}}^+<10$) for which inertial effects begin to dominate; and (iii) the inertial region for ${\rm{St}}^+>10$. Our simulation captures the expected trends across these regimes.

\subsection{Impact dynamics}

\begin{figure}
  \centering
  \begin{subfigure}[t]{0.49\textwidth}
    \centering
    \includegraphics[width=\textwidth]{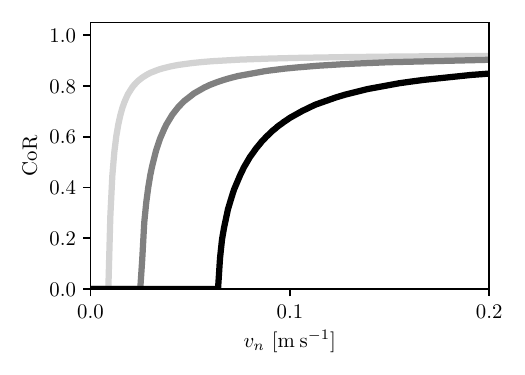}
    \caption{}
    \label{fig:cor}
  \end{subfigure}
  \hfill
  \begin{subfigure}[t]{0.49\textwidth}
    \centering
    \includegraphics[width=\textwidth]{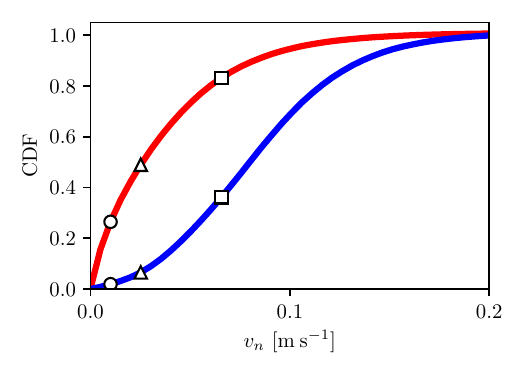}
    \caption{}
    \label{fig:cdf}
  \end{subfigure}

  \caption{Quantification of particle impact dynamics using the (a) coefficient of restitution and (b) normal impact velocity cumulative distribution function. (a) Increasing opacity denotes increasing adhesion number for $\rm{Ad}\in[15, 50, 150]$. (b) ${\rm{St}}^+=10$ (red) and ${\rm{St}}^+=30$ (blue). Symbols in (b) denote the CDF values at the critical sticking velocities as taken from (a); $\rm{Ad}^+=15$ ($\bigcirc$), $\rm{Ad}^+=50$ ($\triangle$), $\rm{Ad}^+=150$ ($\square$).}
  \label{fig:impact}
\end{figure}

Figure~\ref{fig:impact} presents the coefficient of restitution, $\text{CoR} = v_{n}^f / v_{n}^i$, as a function of $\rm{Ad}^+$. The superscripts $f$ and $i$ denote the normal post- and pre-impact velocity, respectively. Also shown are cumulative distribution functions (CDFs) of impact velocity. With our definition of $\rm{Ad}^+$, the CoR is solely a function of the adhesion number. Figure~\ref{fig:cor} shows the non-linear dependence on the rebound velocity as a function of impact velocity. As velocity increases, the CoR tends toward the viscous CoR, $e_0=0.9$. We note that high-velocity events are not present in the turbulent channel. As a result, plastic deformation is unlikely and our choice of visco-elastic adhesion model is deemed appropriate for this flow.

For each $\rm{Ad}^+$, there is a critical velocity below which particles do not rebound ($\text{CoR}=0$). The markers in Fig.~\ref{fig:cdf} denote the points on the CDF corresponding to each critical velocity. The CDFs show that for $\rm{Ad}^+=15$, approximately 25\% of ${\rm{St}}^+=10$ particles stick upon impact; this fraction rises to $\sim50\%$ at $\rm{Ad}^+=50$ and $\sim85\%$ at $\rm{Ad}^+=150$. In contrast, the higher-inertia ${\rm{St}}^+=30$ particles exhibit significantly less sticking, even at high adhesion, with only $\sim40\%$ sticking at $\rm{Ad}^+=150$. Among particles that do rebound, visco-elastic and adhesive losses reduce their rebound energy. For example, $\sim60\%$ of ${\rm{St}}^+=30$ particles impact with normal velocities $\leq \SI{0.1}{\meter\per\second}$, and at $\rm{Ad}^+=150$, these particles rebound with only $\sim70\%$ of that velocity. This energy loss reduces the probability of re-entrainment by near-wall sweep-ejection events, as they possess less energy to rebound to a height at which these events are more energetic.

\subsection{Turbulence and spatial correlation}

\begin{figure}
    \centering
    \includegraphics[width=\linewidth]{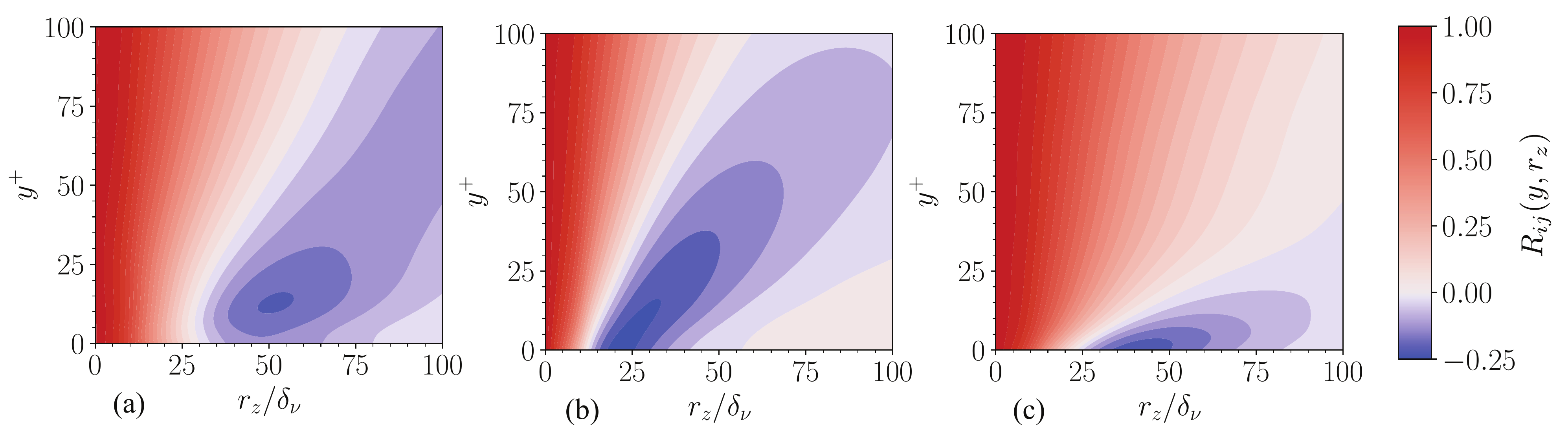}
    \caption{Two-point fluid velocity fluctuation correlations as a function of spanwise separation, $r_z$, and wall-normal distance, $y$, for (a) streamwise velocity, (b) wall-normal velocity, and (c) spanwise velocity.}
    \label{fig:tpVel}
\end{figure}

We now quantify how turbulent structures influence deposition by analyzing two-point velocity correlations. These correlations characterize the spatial structure of turbulent velocity fluctuations that mediate particle transport. The spanwise two-point correlation as a function of wall-normal distance is given by
\begin{equation}
    R_{ij}(y,r_z) = \frac{\mean{u_i'(x,y,z) u_j'(x,y,z+r_z)}_x}{\mean{u_i'(x,y,z)u_j'(x,y,z)}_x},
    \label{eq:tpVel}
\end{equation}
where $r_z$ is the spanwise separation, $u'_i=u_i-\mean{u_i}$ is the fluid velocity fluctuation, $\mean{\cdot}_x$ denotes streamwise and temporal averaging. Correlations are also averaged between the top and bottom halves of the channel. Figure~\ref{fig:tpVel} shows $R_{ij}$ for the streamwise, wall-normal, and spanwise velocities. First, we point to the structure of the distinct regions of negative correlation in each velocity component. Low-speed streaks are clearly identified in the streamwise two-point correlation and are located in $y^+\in[10,15]$ above the wall and spaced approximately $50\delta_\nu$ in the spanwise directions. This spanwise spacing is consistent with previous observations \cite{lagraa2004characterization}. We also note the existence of the strong negative correlation in the wall-normal velocity in $y^+\in(0,15]$, spaced by approximately $25\delta_\nu$, and the on-wall negative correlation in the spanwise velocity spaced between $25\delta_\nu-50\delta_\nu$. Note that these spacings represent half spacings, i.e., the correlations are symmetric about $r_z=0$.

\begin{figure}
    \centering
    \begin{subfigure}[t]{0.49\textwidth}
      \centering
      \includegraphics[width=\textwidth]{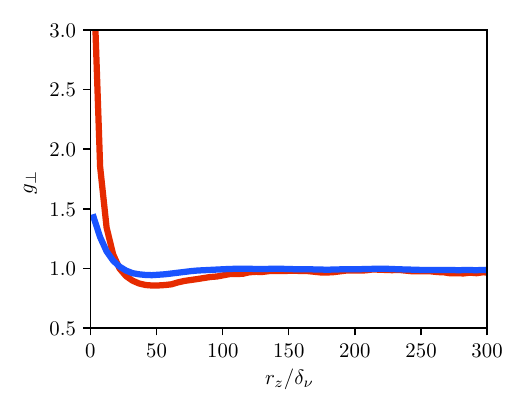}
      \caption{Spanwise RDF.}
    \end{subfigure}
    \hfill
    \begin{subfigure}[t]{0.49\textwidth}
      \centering
      \includegraphics[width=\textwidth]{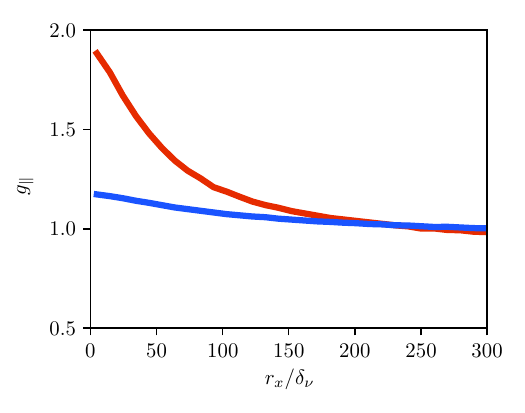}
      \caption{Streamwise RDF.}
    \end{subfigure}
    \caption{One-dimensional radial distribution functions (RDFs) for particles lying in $y^+\in(10, 15)$. ${\rm{St}}^+=10$ (red) and ${\rm{St}}^+=30$ (blue).}
    \label{fig:nearwallRDF}
\end{figure}

Figure~\ref{fig:nearwallRDF} shows one-dimensional streamwise and spanwise radial distribution functions (RDFs) for particles in $y^+\in[10,15]$, the region associated with low-speed streaks as identified in Fig.~\ref{fig:tpVel}. The RDF is defined as
\begin{equation}
    g_{\parallel/\perp} = \frac{N_i/\Delta'_i}{\rho_{\parallel/\perp}},
\end{equation}
where $N_i$ is the number of particle pairs in bin $i$, $\Delta'_i=L/256$ is the bin width, and $\rho_{\parallel/\perp}$ is the expected pair density under a uniform distribution. The near-wall RDFs in Fig.~\ref{fig:nearwallRDF} are computed for particles with purely elastic particle-wall collisions and are averaged in time. Consistent with the observations in \cite{Soldati2009-wl}, the strongest turbulence response is seen for the ${\rm{St}}^+=10$ particles, as shown through the more extreme values in the RDFs compared to the ${\rm{St}}^+=30$ particles. Notably, while the length scales associated with the decay of the streamwise RDF are similar for both Stokes numbers, the shape of the spanwise RDFs differs in both magnitude and correlation length. The ${\rm{St}}^+=10$ particles exhibit stronger clustering for small $r_z$ and the minimum point in the RDF corresponds approximately to the half spacing of the low-speed streaks ($\sim50\delta_\nu$). This is indicative of the $\rm{St}^+=10$ particles responding to the two-point structure of the near-wall turbulence, exhibiting spatial patterns consistent with the coherent structures.

\subsection{Deposit morphology}

\begin{figure}
    \centering
    \includegraphics[width=0.95\linewidth]{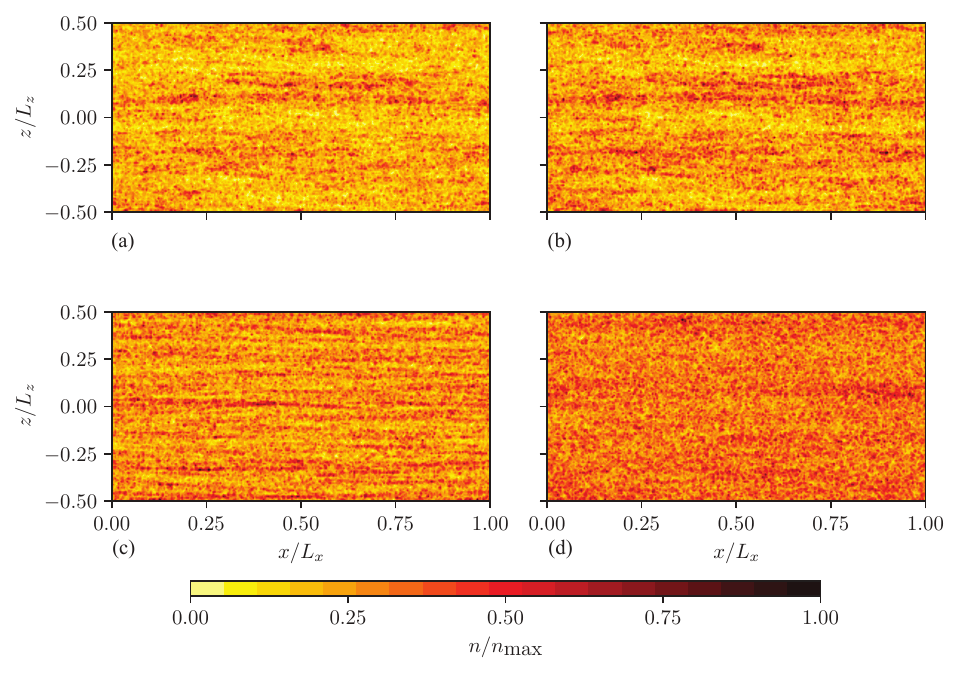}
    \caption{Number density contours for particles deposited on the bottom wall of the channel at $t^+\approx 600$. (a) ${\rm{St}}^+=10$, $\rm{Ad}=150$; (b) ${\rm{St}}^+=10$, $\rm{Ad}=\infty$; (c) ${\rm{St}}^+=30$, $\rm{Ad}=150$; (d) ${\rm{St}}^+=30$, $\rm{Ad}=\infty$.}
    \label{fig:ndenCont}
\end{figure}

Figure~\ref{fig:ndenCont} shows the normalized number density contours for both Stokes numbers with $\rm{Ad}^+=150$ and $\rm{Ad}^+=\infty$. We note that there is little discernible difference in the contours for finite adhesion numbers, hence why only two $\rm{Ad}^+$ are shown. The number density is computed according to
\begin{equation}
    n(\bm{x},t) = \sum_{i=1}^{N_{\rm{dep,b}}}\mathcal{W}(\bm{x}-\bm{x}_{p,i})
\end{equation}
where $N_{\rm{dep,b}}$ is the number of particles deposited on the bottom wall and $\mathcal{W}$ is a bi-linear interpolation weight with units area$^{-1}$. The mesh for $n$ has $N_x\times N_z=256\times 128$ grid points. Contours are normalized by their maxima to isolate deposit geometry from the total number of particles on the wall. 

For ${\rm St}^+ = 10$ particles, the deposit morphology is nearly identical between the finite- and infinite-adhesion cases, consistent with the high sticking probability observed in Fig.~\ref{fig:impact}. In contrast, ${\rm St}^+ = 30$ particles exhibit pronounced streaking patterns only when adhesion is finite. The all-stick case yields a more uniform deposition pattern, despite identical initial impact locations. This indicates that post-collision motion--rather than initial impact--is the dominant contributor to deposition heterogeneity at ${\rm St}^+ = 30$. Particle rolling and sliding play a significant role in shaping the final deposit morphology. This behavior is likely to be even more important in cases where inter-particle collisions are considered, as deposited particles can interact and form agglomerates on the surface.

Figures~\ref{fig:nearwallRDF} and~\ref{fig:ndenCont} illustrate the relationship between wall deposition patterns and particle clustering just above the wall. Particles with ${\rm St}^+ = 10$ exhibit significantly more clustering prior to wall impact. This is evident when comparing the all-stick cases: the initial impact points for ${\rm St}^+ = 10$ show more pronounced streaks and voids than those for ${\rm St}^+ = 30$. We also highlight the presence of ``hot spot" deposition patterns in the ${\rm St}^+ = 10$ case. While \citet{narayanan2003} attributed streak-like deposition to particles diffusing toward the wall from elongated streaks in the viscous sublayer, our results suggest an alternative mechanism. Specifically, the hot spots observed for ${\rm St}^+ = 10$ appear to result from clustered particles impacting the wall simultaneously. This indicates that deposition in this regime is driven by sweep events in which particles arrive at the wall already clustered. Such behavior is consistent with the findings of \citet{marchioli2002}, who showed that sweep events entrain and transport particles toward the wall.

\begin{figure}
    \centering
    \begin{subfigure}{0.49\linewidth}
       \centering
       \includegraphics[width=\linewidth]{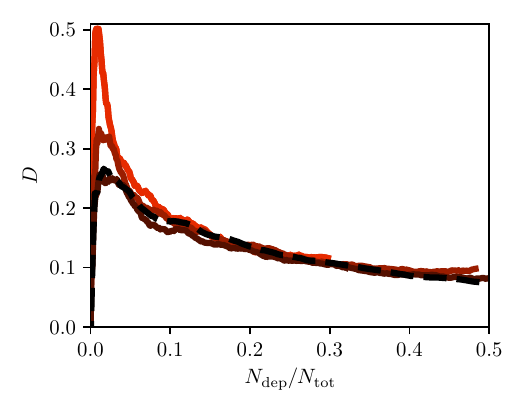}
       \caption{${\rm{St}}^+=10$}
    \end{subfigure}
    \begin{subfigure}{0.49\linewidth}
       \centering
       \includegraphics[width=\linewidth]{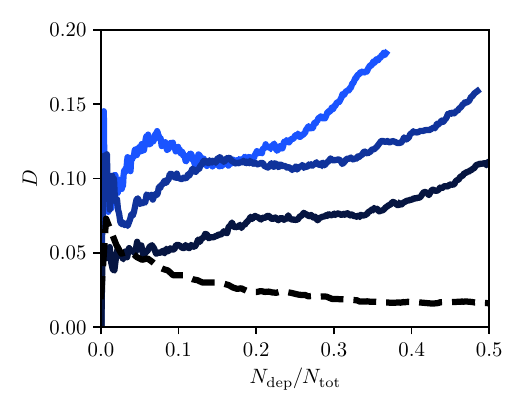}
       \caption{${\rm{St}}^+=30$}
    \end{subfigure}
    \caption{Comparison of the clustering parameter \eqref{eq:Dstat} across all cases plotted as a function of the fraction of particles on the walls. $N_{\rm{dep}}$ is the total number of deposited particles. Increasing line darkness corresponds to increasing adhesion number for $\rm{Ad}^+\in[15, 50, 150, \infty]$. $\rm{Ad}^+=\infty$ (--\,--).}
    \label{fig:Dstat}
\end{figure}

To quantify heterogeneity over time, we consider a clustering parameter, given by \cite{eatonPreferentialConcentrationParticles1994}
\begin{equation}
    D=\frac{\sigma - \sigma_p}{\lambda},
    \label{eq:Dstat}
\end{equation}
where $\sigma$ is the standard deviation of the observed number density, $\sigma_{p}$ is that of a Poisson distribution, and $\lambda = \mean{N_{p}}$. Higher values of $D$ indicate greater heterogeneity. Figure~\ref{fig:Dstat} shows this clustering metric for two Stokes numbers and all $\rm{Ad}^+$ plotted against the fraction of particles on the wall. The values of $D$ are averaged between deposits on the top and bottom walls. The trends suggested by the contours in Fig.~\ref{fig:ndenCont} are again represented in Fig.~\ref{fig:Dstat}: the ${\rm{St}}^+=10$ particles show initially heterogeneous deposition patterns that trend towards homogeneity as particles continue to deposit. Furthermore, the ${\rm{St}}^+=30$ particles with finite adhesion show increasing heterogeneity with time, with lower adhesion numbers promoting higher levels of heterogeneity. We attribute this evolution towards a more clustered state to the larger Stokes number particles responding to turbulent structures near the wall and overcoming adhesive rolling resistance. The disparity in the all-stick results against the finite adhesion results for the ${\rm{St}}^+=30$ particles supports our claim of on-wall particle motion being a significant source of the heterogeneous patterns seen in Fig.~\ref{fig:ndenCont} (i.e., initially homogeneous deposits tend toward heterogeneity).

\begin{figure}
    \centering
    \begin{subfigure}[t]{0.49\textwidth}
      \centering
      \includegraphics[width=\textwidth]{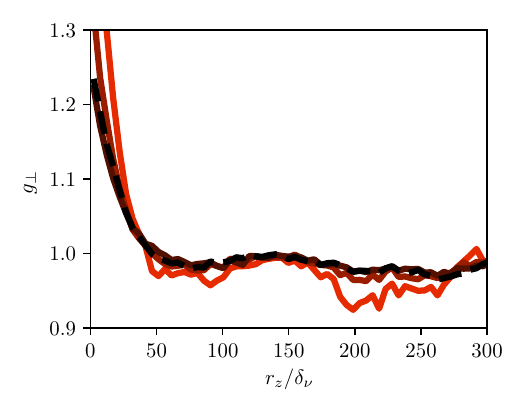}
      \caption{${\rm{St}}^+=10$, $t^+\approx 600$.}
    \end{subfigure}
    \hfill
    \begin{subfigure}[t]{0.49\textwidth}
      \centering
      \includegraphics[width=\textwidth]{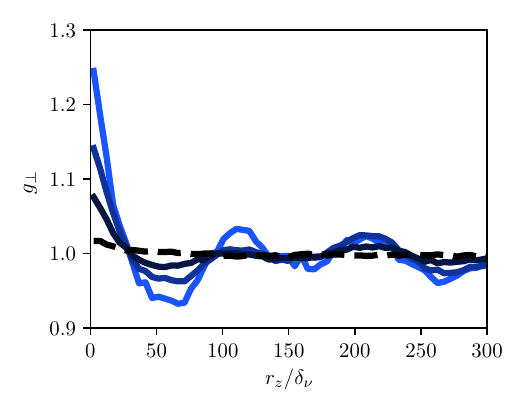}
      \caption{${\rm{St}}^+=30$, $t^+\approx 600$.}
    \end{subfigure}
    \hfill
    \begin{subfigure}[t]{0.49\textwidth}
      \centering
      \includegraphics[width=\textwidth]{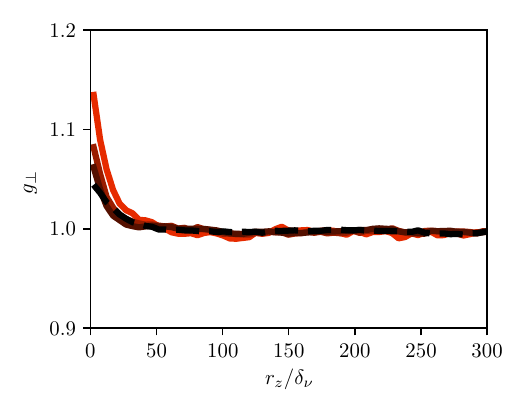}
      \caption{${\rm{St}}^+=10$, $t^+\approx 1200$.}
    \end{subfigure}
    \hfill
    \begin{subfigure}[t]{0.49\textwidth}
      \centering
      \includegraphics[width=\textwidth]{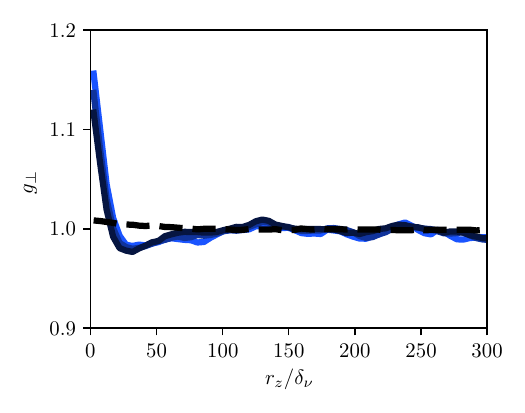}
      \caption{${\rm{St}}^+=30$, $t^+\approx 1200$.}
    \end{subfigure}
    \caption{One-dimensional spanwise radial distribution functions computed from deposited particles at varying times, adhesion numbers, and Stokes number. Increasing darkness corresponds to increasing adhesion number for $\rm{Ad}^+\in[15, 50, 150, \infty]$. $\rm{Ad}^+=\infty$ (--\,--).}
    \label{fig:spanwiseRDFs}
\end{figure}

\begin{figure}
    \centering
    \begin{subfigure}[t]{0.49\textwidth}
      \centering
      \includegraphics[width=\textwidth]{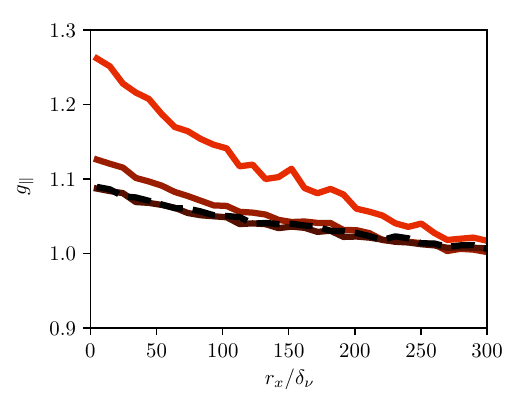}
      \caption{${\rm{St}}^+=10$, $t^+\approx 600$}
    \end{subfigure}
    \hfill
    \begin{subfigure}[t]{0.49\textwidth}
      \centering
      \includegraphics[width=\textwidth]{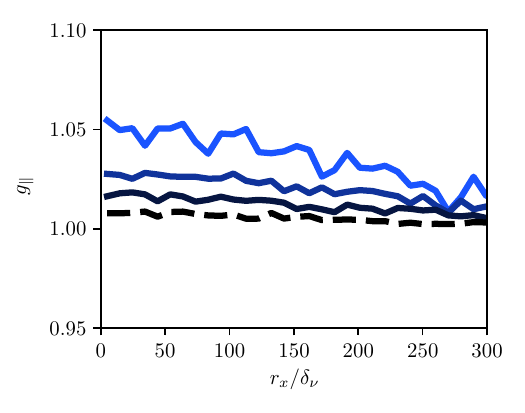}
      \caption{${\rm{St}}^+=30$, $t^+\approx 600$}
    \end{subfigure}
    \hfill
    \begin{subfigure}[t]{0.49\textwidth}
      \centering
      \includegraphics[width=\textwidth]{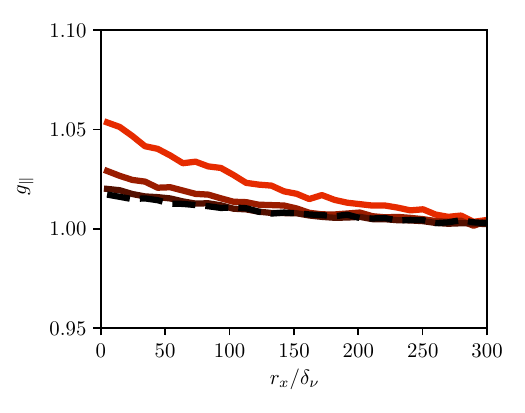}
      \caption{${\rm{St}}^+=10$, $t^+\approx 1200$}
    \end{subfigure}
    \hfill
    \begin{subfigure}[t]{0.49\textwidth}
      \centering
      \includegraphics[width=\textwidth]{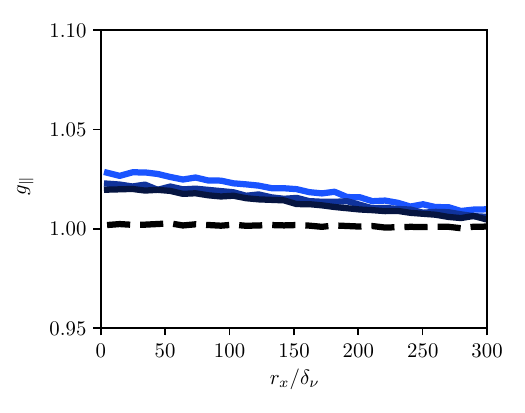}
      \caption{${\rm{St}}^+=30$, $t^+\approx 1200$}
    \end{subfigure}
    \caption{One-dimensional streamwise radial distribution functions computed from deposited particles at varying times, adhesion numbers, and Stokes number. Increasing darkness corresponds to increasing adhesion number for $\rm{Ad}\in[15, 50, 150, \infty]$. $\rm{Ad}^+=\infty$ (--\,--).}
    \label{fig:streamwiseRDFs}
\end{figure}

Figures \ref{fig:spanwiseRDFs} and \ref{fig:streamwiseRDFs} compare instantaneous spanwise and streamwise RDFs for all cases computed from the particles that are in contact with the wall. Averaging is performed between the top and bottom wall. The RDFs are presented at two different times to expose the temporal evolution of the deposits. It is again apparent from Fig.~\ref{fig:spanwiseRDFs} that the ${\rm{St}}^+=30$ particles with finite adhesion exhibit more heterogeneous deposits than the ${\rm{St}}^+=10$ particles, as indicated by the void ($g<1$) regions in the ${\rm{St}}^+=30$ spanwise RDFs. Though the higher values in the RDF for the ${\rm{St}}^+=10$ particles at $t^+=600$ indicate more dense clusters, the void regions for the higher Stokes number particles indicate streak-dominant deposit morphology. Furthermore, similar to the ${\rm{St}}^+=10$ spanwise near-wall RDF in Fig.~\ref{fig:nearwallRDF}, the ${\rm{St}}^+=30$ spanwise RDF computed on the wall in Fig.~\ref{fig:spanwiseRDFs} shows a period of oscillation that corresponds approximately to the spacing between the low-speed streaks in Fig.~\ref{fig:tpVel}. We also turn back to the earlier discussion on ${\rm{St}}^+=10$ particles being deposited in clusters. The RDFs indicate that the ${\rm{St}}^+=10$ particles deposits contain many particles, as shown by the larger values of $g_\perp$ at early times, the lack of a void region in $g_\perp$, and the steeper decay in $g_\parallel$ when compared to $g_\parallel$ in ${\rm{St}}^+=30$ case. These low aspect ratio deposits are again indicative of ${\rm{St}}^+=10$ particles depositing together in coherent events. Sw-eps-converted-to.pdf that approach the wall carry ${\rm{St}}^+=10$ particles, the sw-eps-converted-to.pdf widen as they approach the wall and lift back up, and the ${\rm{St}}^+=10$ particles are ejected and deposit on the wall.

Figures \ref{fig:spanwiseRDFs} and \ref{fig:streamwiseRDFs} also show a general trend with finite adhesion numbers, with lower adhesion numbers displaying more heterogeneity than higher adhesion numbers. The disparity is more present at early times, with the later time results showing near-collapse in the spanwise and streamwise RDFs. We also note the late time behavior for the ${\rm{St}}^+=30$ spanwise RDFs, namely the movement of the minimum value of the RDF to smaller $r_z$. This collapse towards smaller streak spacing corresponds to the on-wall wall-normal and spanwise velocity correlations seen in Fig.~\ref{fig:tpVel}. This is consistent with the mechanisms for ``trapped'' particles as determined by \citet{marchioli2002}. They point to vortex generation at the wall as the source of strong spanwise velocities which concentrate particles in very near-wall low-speed, high-strain regions. This drives particles to concentrate in the region of zero spanwise correlation, or at $\sim25\delta_\nu$ as identified in Fig.~\ref{fig:tpVel}.

\begin{figure}[hbt]
    \centering
    \includegraphics[width=0.33\linewidth]{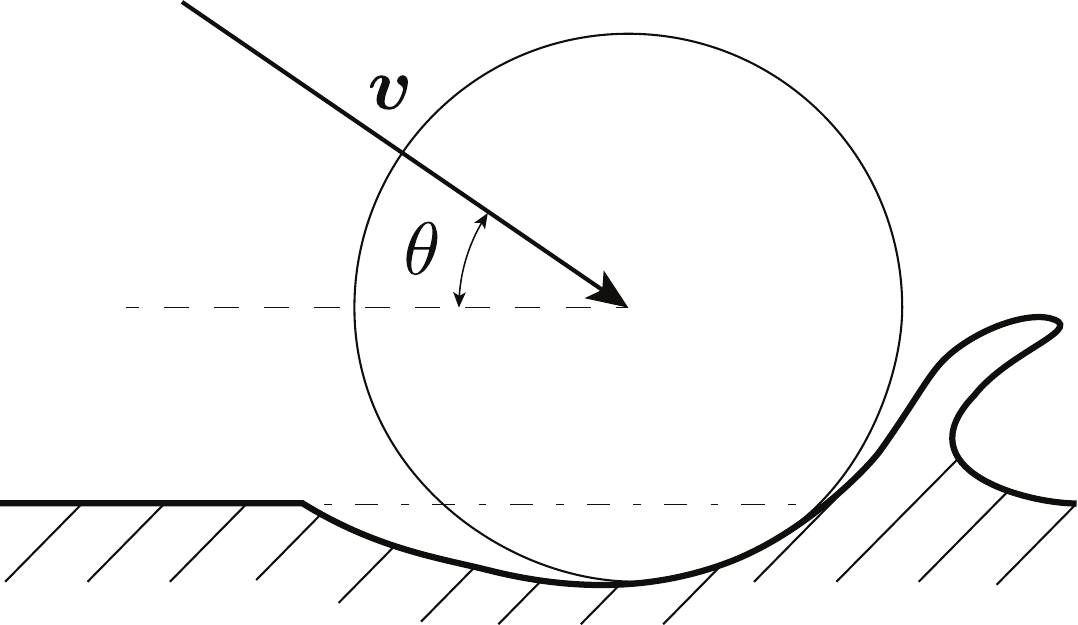}
    \caption{A schematic depicting abrasive wear. $\theta$ is the impact angle and $v=||\bm{v}||$ is the magnitude of the impact velocity.}
    \label{fig:wearPart}
\end{figure}

\subsection{Wear}
Particle impingement can lead to material erosion, providing a means to quantify the consequences of heterogeneous deposition. To estimate erosion, we employ the Finnie model~\cite{finnie1960erosion}, which predicts the eroded volume $W$ due to abrasive wear as
\begin{equation}
    W=\frac{c}{\phi}\frac{m||\bm{v}||^2}{2}\frac{1}{P}f(\theta),
\end{equation}
where $\theta$ is the impact angle and $P$ is the plastic flow pressure (see Fig.~\ref{fig:wearPart}). Following \citet{Zhao2017-wm}, we rewrite this for the soft-sphere case of sustained particle-wall contact as $W=E_{\textrm{Shear}}/P$, where
\begin{equation}
    E_{\textrm{Shear}}=-\int_{t_0}^{t_1} \bm{F}_t \cdot \bm{v}_r \dd{t} \quad \textrm{when} \quad \bm{F}_t\cdot\bm{v}_r<0 \textrm{ and } \bm{F}_t\cdot\bm{v}<0,
    \label{eq:wear}
\end{equation}
where $\bm{F}_t$ is the tangential collision force and $\bm{v}_r$ is the relative contact velocity.

\begin{figure}
    \centering
    \includegraphics[width=0.95\linewidth]{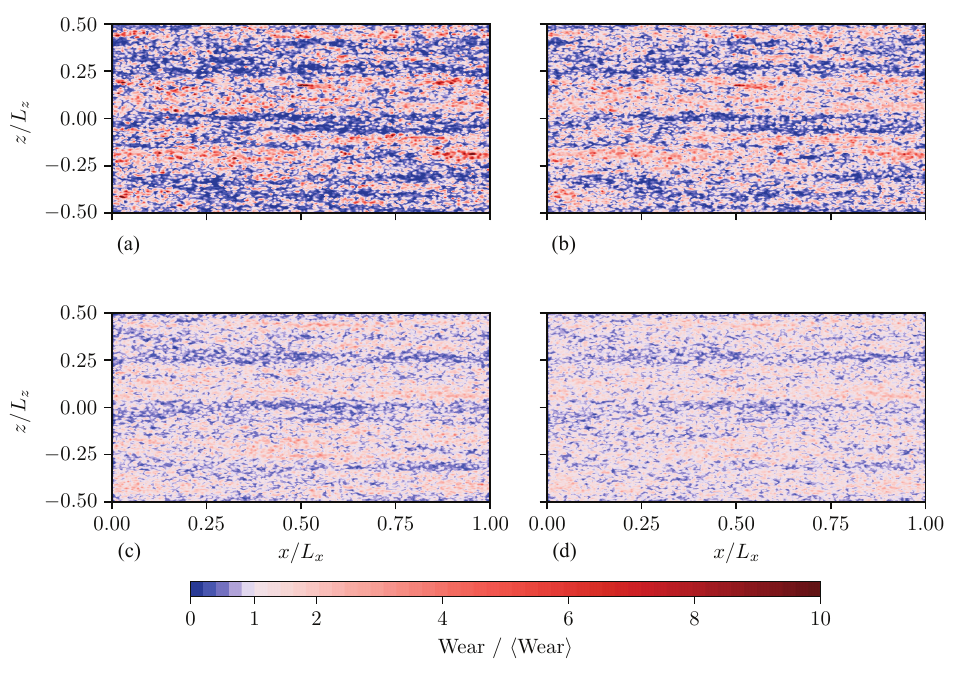}
    \caption{Wear contours at $t^+\approx600$ for (a) ${\rm{St}}^+=10$, $\rm{Ad}=15$; (b) ${\rm{St}}^+=10$, $\rm{Ad}=150$; (c) ${\rm{St}}^+=30$, $\rm{Ad}=15$; and (d) ${\rm{St}}^+=30$, $\rm{Ad}=150$ particles. Wear is normalized by the mean value of each case.}
    \label{fig:wearCont}
\end{figure}

Figure~\ref{fig:wearCont} presents wear contours for both Stokes numbers at low and high $\rm{Ad}^+$. Unlike the number density patterns in Fig.~\ref{fig:ndenCont}, the wear distribution is more heterogeneous for ${\rm St}^+ = 10$ particles. In this case, localized regions exhibit wear rates exceeding ten times the mean. For ${\rm St}^+ = 30$, wear is more uniformly distributed, with peak values reaching only about three times the mean. This difference reflects the fact that surface erosion is primarily driven by particle kinetic energy at the moment of initial impact. Higher-inertia (${ \rm St}^+ = 30$) particles exhibit more ``free-flight" behavior \cite{narayanan2003}, spreading their impact energy more uniformly across the channel width. Adhesion has a secondary effect: larger $\rm{Ad}^+$ values lead to slightly reduced heterogeneity. Since abrasive wear is driven largely by the transfer of kinetic energy upon impact--and this energy is unaffected by adhesion in our simulations--the role of adhesion is limited.

We note that in some extreme cases, low-$\rm{Ad}^+$ particles may temporarily adhere to the wall, then resuspend and impact again. However, these secondary impacts are typically associated with less energetic near-wall motions and are unlikely to contribute significantly to wear. Instead, the highest wear potential arises from particles carried by energetic sweep events that deliver them to the wall with sufficient momentum.

\begin{figure}
    \centering
    \begin{subfigure}[t]{0.49\textwidth}
      \centering
      \includegraphics[width=\textwidth]{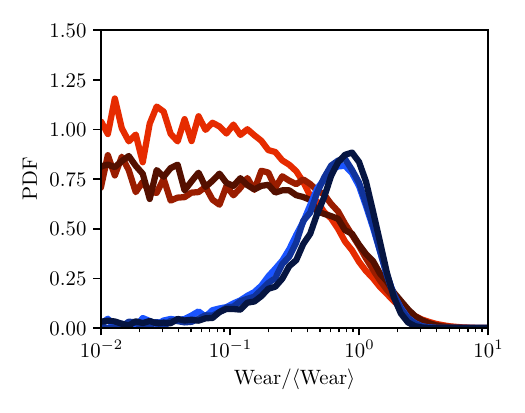}
      \caption{}
    \end{subfigure}
    \hfill
    \begin{subfigure}[t]{0.49\textwidth}
      \centering
      \includegraphics[width=\textwidth]{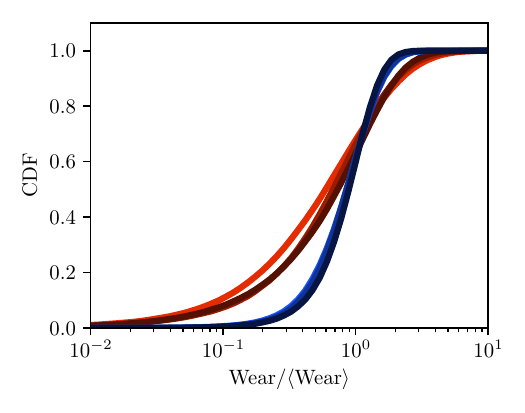}
      \caption{}
    \end{subfigure}
    \caption{Comparisons of (a) probability density functions and (b) cumulative distribution functions for normalized wear. ${\rm{St}}^+=10$ (red), ${\rm{St}}^+=30$ (blue), and increasing darkness corresponds to increasing adhesion number.}
    \label{fig:wearDist}
\end{figure}

Figure~\ref{fig:wearDist} shows probability density functions and cumulative distribution functions for wear normalized by their mean values. In this form, distributions that are skewed away from unity are associated with higher levels of heterogeneity. It is apparent from Fig.~\ref{fig:wearDist} that the ${\rm{St}}^+=10$ particles present more heterogeneous wear patterns, with lower adhesion prompting more heterogeneity. These trends with wear and Stokes number further confirm that on-wall heterogeneity is a product of free-stream particle-turbulence interactions. The near-wall clustering of the ${\rm{St}}^+=10$ particles (see Fig.~\ref{fig:nearwallRDF}) is imprinted on the wall. Higher Stokes number particles do not demonstrate the same extent of free-stream concentration, and hence imprint with more uniform wear distributions. 

\section{Summary and conclusion}\label{sec:conclusion}
This study quantified the spatial structure of particle deposition in turbulent channel flow using direct numerical simulation and a visco-elastic soft-sphere collision model with van der Waals adhesion. We focused on how near-wall turbulence, particle inertia, and adhesion jointly influence deposit morphology and surface wear. Particles with ${\rm{St}}^+=10$ preferentially cluster in the turbulence and exhibit heterogeneous impact patterns, indicative of deposition by sweep events. The ${\rm{St}}^+=30$ particles deposit more uniformly upon initial impact but show significant post-collision motion, leading to streak-like deposit structures when adhesion is finite. Two-point velocity correlations and radial distribution functions revealed that deposit heterogeneity correlates with low-speed streak spacing ($\sim50\delta_\nu$) and near-wall turbulent structures and events. Finally, energy-based wear modeling shows that this heterogeneity is imprinted on the surface through abrasive wear patterns, particularly for ${\rm{St}}^+\sim10$ particles, whose clustered impact locations lead to hot spots of wear that exceed ten times the mean. These results underscore the importance of resolving particle-turbulence interactions and modeling particle sliding and rolling to accurately predict deposition and wear. By connecting near-wall turbulence statistics to on-wall deposit morphology, this work provides a framework to better understand and predict particulate fouling and erosion in high-temperature, confined systems such as gas turbine engines.  

Although Reynolds-averaged Navier--Stokes (RANS) and, more recently, large-eddy simulations (LES), represent the workhorses for turbomachinery flow modeling, current subgrid-scale models rely on one-point turbulence statistics that are inadequate for capturing the mechanisms driving heterogeneous particle deposition and wear, as identified in this work. We therefore advocate for the development of accurate and computationally efficient subgrid-scale models that explicitly incorporate two-point turbulence statistics to better represent particle dispersion and clustering dynamics.

\section{Acknowledgments}
This research was supported by the Office of Naval Research under Award No. N00014-23-1-2369 and by the National Science Foundation under NSF CBET-2035489. The computational resources were provided by Advanced Research Computing at the University of Michigan, Ann Arbor.

\appendix
\section{Soft-sphere model validation}\label{sec:appendix}

\begin{figure}
    \centering
    \includegraphics[width=0.5\linewidth]{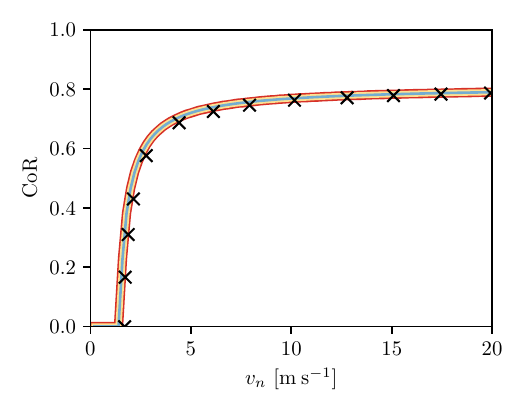}
    \caption{Validation of the soft-sphere model and artificial softening scheme from \cite{Haervig2017-pb} used in this study. Experiments are by \citet{wallMeasurementsKineticEnergy1990} ($\times$). $E/E_{\rm{mod}}=1$ (red), $E/E_{\rm{mod}}=10$ (yellow), and $E/E_{\rm{mod}}=1,000$ (blue).}
    \label{fig:wallValid}
\end{figure}

The soft-sphere model in Sec. \ref{sec:ssmodel} was validated using a single-particle rebound case. The case is setup such that an individual particle is initialized above a wall at some prescribed position and velocity, as specified by the user. No near-wall time step modification or sub-step scheme is used; there is no fluid solve and particles are assumed to move through a fluid that has zero velocity. Drag and lift forces are not included. Figure~\ref{fig:wallValid} shows excellent agreement with experimental data, with no change as the softening ratio $E/E_{\rm{mod}}$ increases. The material properties were taken from \cite{marshallAdhesiveParticleFlow2014, wallMeasurementsKineticEnergy1990}. Particles are ammonium fluorescein impacting a silicon target, consistent with the experiments by \citet{wallMeasurementsKineticEnergy1990}. 

\bibliography{adh}

\end{document}